\documentclass[11pt]{article}
\usepackage[intlimits]{amsmath}
\usepackage{amssymb}
\usepackage{psfrag}
\usepackage{graphicx}

\def\be{\begin{equation}}
\def\ee{\end{equation}}
\def\bea{\begin{eqnarray}}
\def\eea{\end{eqnarray}}

\def\mbf#1{\mathchoice{\hbox{\boldmath $\displaystyle #1$}}
        {\hbox{\boldmath $\textstyle #1$}}{\hbox{\boldmath $\scriptstyle #1$}}
        {\hbox{\boldmath $\scriptscriptstyle #1$}}}

\begin{document}

\begin{titlepage}
\begin{flushright}
ECT$^*$-06-18\\
HD-THEP-06-29\\
\end{flushright}
\vfill
\begin{center}
\boldmath
{\LARGE{\bf Bounds on Ratios of DIS Structure Functions}}\\[.2cm]
{\LARGE{\bf from the Color Dipole Picture}}
\unboldmath
\end{center}
\vspace{1.2cm}
\begin{center}
{\bf \Large
Carlo Ewerz\,$^{a,1}$, Otto Nachtmann\,$^{b,2}$
}
\end{center}
\vspace{.2cm}
\begin{center}
$^a$
{\sl
ECT\,$^*$, Strada delle Tabarelle 286, 
I-38050 Villazzano (Trento), Italy}
\\[.5cm]
$^b$
{\sl
Institut f\"ur Theoretische Physik, Universit\"at Heidelberg\\
Philosophenweg 16, D-69120 Heidelberg, Germany}
\end{center}                                                                
\vfill
\begin{abstract}
\noindent
We derive bounds on ratios of deep inelastic nucleon structure functions 
from the color dipole picture of high energy photon-hadron scattering. 
We find an upper bound on the ratio $R=\sigma_L/\sigma_T$ 
of the total cross sections for longitudinally and transversely polarized 
photons. We further obtain bounds on the ratio of deep inelastic 
structure functions $F_2$ taken at the same energy but 
at different photon virtualities. 
It is shown that these bounds can be used to constrain the range of 
applicability of the dipole picture. 
\vfill
\end{abstract}
\vspace{5em}
\hrule width 5.cm
\vspace*{.5em}
{\small \noindent
$^1$ email: Ewerz@ect.it \\
$^2$ email: O.Nachtmann@thphys.uni-heidelberg.de
}
\end{titlepage}

\section{Introduction}

The precise determination of the 
proton structure function $F_2$ 
in deep inelastic scattering (DIS) at high energy has been a major  
achievement of the H1 and ZEUS experiments at the HERA collider 
\cite{Breitweg:2000yn,Adloff:2000qk,Chekanov:2001qu,Adloff:2003uh,Chekanov:2003yv}. 
Much attention has been given to the region of 
small Bjorken-$x$ where $F_2$ exhibits a significant growth 
as $x$ decreases. This kinematical region is very interesting 
since here one can study the properties of a very dense system of partons. 
At ever smaller values of $x$ or with increasing energy  
the parton densities should become so large 
that parton recombination processes are significant. It is expected 
that then a saturation of parton densities takes place, eventually 
taming the further rise of $F_2$ 
\cite{Gribov:1984tu,Mueller:1985wy,Mueller:1989st}. 
To date it is still an open question whether the energy available at 
HERA is already sufficiently high to probe this interesting and so 
far unexplored regime of QCD. 

The smallest values of $x$ accessible at HERA occur for relatively small 
photon virtualities $Q^2$, thus prohibiting the use of perturbative QCD. 
In addition, also the high parton densities in this region make the 
theoretical description rather involved. 
In order to study possible saturation effects one 
therefore has to use theoretical models which do or do not incorporate  
saturation effects and to compare them with the experimental data. 
A prominent example of a saturation model is the 
Golec-Biernat-W\"usthoff model 
\cite{Golec-Biernat:1998js,Golec-Biernat:1999qd}. 
The most widely used framework for implementing such models is 
the color dipole picture of high energy scattering 
\cite{Nikolaev:1990ja,Nikolaev:et,Mueller:1993rr}. 
The dipole picture is motivated by perturbation theory and 
describes photon-proton scattering as a two-step process. 
In the first step the photon splits into a quark-antiquark 
pair -- the color dipole. In the second step this quark-antiquark 
pair scatters off the proton in the forward direction. 
While the dipole picture is certainly 
accurate at large photon virtualities and asymptotically large 
energy, the situation is less clear at moderate or low $Q^2$, 
and for presently available energies. 
To obtain the dipole picture from a genuinely nonperturbative 
description of photon-proton scattering requires a number of 
approximations and assumptions \cite{Ewerz:2004vf,Ewerz:2006vd}. 
It is intrinsically difficult to determine their accuracy and 
to estimate the size of potential corrections to the dipole picture. 
Therefore it is important to constrain the range of applicability 
of the dipole picture before one can use it to draw conclusions about 
saturation, for example. 

In this Letter we present bounds on ratios of structure functions 
from the dipole picture. These bounds are derived only from 
the general formulae constituting the dipole picture 
and are independent of any particular model for the 
dipole-proton scattering process. Any violation of these bounds 
by the experimental data would indicate a breakdown of the 
dipole picture. We show that in this way our bounds can indeed 
be used to constrain the range of applicability of the dipole picture. 

\section{The Dipole Picture}

The amplitude for the splitting of a transversely (T) or 
longitudinally (L) polarized photon into a quark-antiquark 
pair of given flavor $q$ in a high energy photon-hadron scattering 
process is given by the so-called photon wave function $\psi^{(q)}_{T,L}$. 
It depends on the photon virtuality $Q^2$ and on the quantum 
numbers specifying the color dipole, which are its  
size and orientation in the transverse space of the scattering process 
described by a two-dimensional vector $\mbf{r}$, its 
longitudinal momentum $\mbf{q}$ in a given reference frame, 
the momentum fractions $\alpha$ and $(1-\alpha)$ of $\mbf{q}$ 
carried by the quark and antiquark, respectively, and finally the 
spin orientations of the quark and antiquark. In the following we will 
always sum over these spin orientations. The square of the 
photon wave function is obtained in leading order in the 
coupling constants $\alpha_{\rm em}$ and $\alpha_{\rm s}$ as 
\be
\label{sumpsi+dens}
\left| \psi_T^{(q)} (\alpha, \mbf{r},Q) \right|^2 
= 
\frac{3}{2 \pi^2} \, \alpha_{\rm em} Q_q^2 
\left\{ \left[ \alpha^2 + (1-\alpha)^2 \right] 
\epsilon_q^2 [K_1(\epsilon_q r) ]^2 
+ m_q^2 [K_0(\epsilon_q r) ]^2 
\right\} 
\ee
and 
\be
\label{sumpsiLdens}
\left|\psi_L^{(q)}(\alpha, \mbf{r},Q) \right|^2 
=
\frac{6}{\pi^2} \, \alpha_{\rm em} Q_q^2 
Q^2 [\alpha (1-\alpha)]^2 [K_0(\epsilon_q r) ]^2 
\ee
for transversely and longitudinally polarized photons, respectively. 
Here $r=\sqrt{\mbf{r}^2}$, $Q_q$ are the quark charges in 
units of the proton charge, 
and $K_0$ and $K_1$ are modified Bessel functions. 
The quantity 
$\epsilon_q = \sqrt{\alpha (1-\alpha) Q^2 +m_q^2}$ involves the 
quark mass $m_q$. We then define a density for the photon 
wave function by integrating over the longitudinal momentum 
fraction $\alpha$, 
\be
\label{wispsi2}
w^{(q)}_{T,L}(r,Q^2) =
\int^1_0 d\alpha \,
\left|
\psi^{(q)}_{T,L}(\alpha,\mbf{r},Q)
\right|^2 
\,. 
\ee
It describes the probability that a photon of virtuality $Q^2$ 
splits into a color dipole of size $r$ and flavor $q$. 

The second step of the photon-proton scattering process in 
the dipole picture is then the scattering of the dipole of size 
$r$ off the proton, given by the so-called dipole cross section 
$\hat{\sigma}^{(q)}$. It depends on the dipole 
size $r$ and on the squared center-of-mass energy $W^2$ in 
the dipole-proton system. 
Accordingly, the cross section for the photon-proton scattering process 
is then obtained by folding the photon density with the dipole-proton 
cross section, and by summing over all quark flavors $q$, 
\be
\label{sigTLdip}
\sigma_{T,L}(W^2,Q^2)=
\sum_q\int d^2 r \,
w^{(q)}_{T,L}(r,Q^2)\,
\hat{\sigma}^{(q)}(r,W^2) 
\,.
\ee
The integration is 
over all dipole sizes and orientations. For a detailed discussion 
of the dipole picture, its nonperturbative foundations and 
potential correction terms see 
\cite{Ewerz:2004vf,Ewerz:2006vd}. 

Note that the energy variable in the dipole cross section is 
$W^2$. In practical applications of the dipole picture one 
frequently uses Bjorken's $x=Q^2/(W^2+Q^2-m_p^2)$ instead. 
Strictly speaking, this is incorrect, since $Q^2$ is not 
uniquely determined by the properties of the dipole. The use of $x$ 
is motivated by the fact that the photon densities $w_{T,L}$ 
are behaved in such a way that the integral in (\ref{sigTLdip}) 
is dominated by values of $r$ around $4/Q^2$. 
Concentrating on the dominant values of $r$ one can 
therefore trade $W^2$ for $x$. 
For a more detailed discussion of this point see \cite{Ewerz:2006vd}. 
We further note that in the limit of asymptotically high energy $W$ 
the quark and antiquark forming the dipole are on their 
mass shell. In the dipole picture one therefore regards the 
dipole-proton scattering as a physical process. 
As such its cross section $\hat{\sigma}^{(q)}$ has to be 
non-negative. 

Finally, we have for the structure function $F_2$ for $x \ll 1$ 
and $W^2 \gg m_p^2$ 
\be
\label{Ffromsigma}
F_2(W^2,Q^2) = 
\frac{Q^2}{4 \pi^2 \alpha_{\rm em}}  
\left[ \sigma_T(W^2,Q^2) + \sigma_L(W^2,Q^2) \right]
\,,
\ee
and the two terms in this sum are the transverse and 
longitudinal structure functions $F_T$ and 
$F_L$, respectively. 

\boldmath
\section{Bound on $R =\sigma_L/\sigma_T$} 
\unboldmath

We first consider the ratio of the cross sections for 
longitudinally and transversely polarized photons, 
$R(W^2,Q^2) = \sigma_L(W^2,Q^2)/\sigma_T(W^2,Q^2)$, 
or, equivalently, $R=F_L/F_T$. Recalling that $\sigma_L$ and 
$\sigma_T$ in (\ref{sigTLdip}) involve the same factor 
$\hat{\sigma}^{(q)}$, and that this factor as well as the 
photon densities $w^{(q)}_{T,L}$ of (\ref{wispsi2}) are non-negative, 
it is straightforward to obtain upper and lower bounds on $R$ 
in the dipole picture, valid for all $W^2$, 
\be
\label{VI55l}
\min_{q,r} \,
\frac{w^{(q)}_L(r,Q^2)}{w^{(q)}_T(r,Q^2)}
\, \leq \, R(W^2,Q^2)
\, \leq \,\max_{q,r} \,
\frac{w^{(q)}_L(r,Q^2)}{w^{(q)}_T(r,Q^2)} 
\,.
\ee
Here the minimum and maximum are taken over all 
dipole sizes and over all quark flavors.  Note that this bound 
is independent of the choice of energy variable ($W^2$ or $x$) 
in the dipole cross section. To prove (\ref{VI55l}) we 
start from the obvious inequalities 
\be
\min_{q,r} \frac{w_L^{(q)}(r,Q^2)}{w_T^{(q)}(r,Q^2)} 
\, \le \, \frac{w_L^{(q)}(r,Q^2)}{w_T^{(q)}(r,Q^2)}
\, \le \, \max_{q,r} \frac{w_L^{(q)}(r,Q^2)}{w_T^{(q)}(r,Q^2)} 
\,.
\ee
Multiplying by the nonnegative factor $\hat{\sigma}^{(q)} w_T^{(q)}$, 
integrating over $r$ and summing over $q$ leads to inequalities  
equivalent to (\ref{VI55l}). 

Using the explicit formulae 
(\ref{sumpsi+dens}) and (\ref{sumpsiLdens}) we 
find that the lower bound is trivial ($R \ge 0$). 
The most conservative upper bound, valid for 
all $W^2$ and $Q^2$, is obtained by assuming the 
light quarks ($u$, $d$, $s$) to be massless, leading to the 
numerical value 
\be
\label{boundonR}
R(W^2,Q^2) \le 0.37248
\,.
\ee

Similarly, we can consider the cross sections 
$\sigma_{L/T}^{(q)}$ for heavy flavor production, 
$q=c,b$, which are naturally 
obtained via the flavor decomposition of (\ref{sigTLdip}). 
For the ratio $R_q=\sigma_{L}^{(q)}/\sigma_{T}^{(q)}$ 
we derive the bounds 
\be
\label{boundRheavy}
\min_{r} \,
\frac{w^{(q)}_L(r,Q^2)}{w^{(q)}_T(r,Q^2)}
\, \leq \, R_q(W^2,Q^2)
\, \leq \,\max_{r} \,
\frac{w^{(q)}_L(r,Q^2)}{w^{(q)}_T(r,Q^2)} \,,
\ee
valid for all $W^2$. The lower bound on $R_c$ and $R_b$ is again trivial, 
while the upper bound becomes $Q^2$-dependent for heavy quarks. 
\begin{figure}[ht]
\begin{center}
\includegraphics[width=11.4cm]{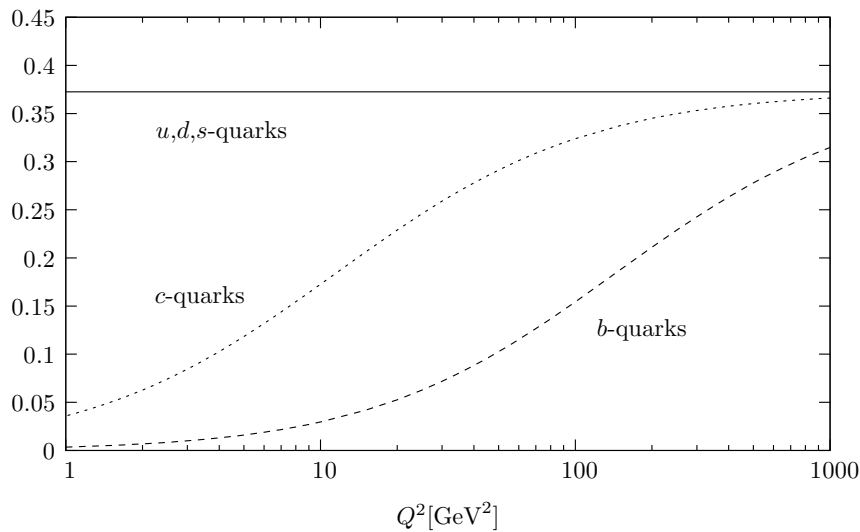}
\end{center}
\caption{The upper bound on $R=\sigma_L/\sigma_T$ 
resulting from the dipole picture as a function of $Q^2$. 
The solid line is for light quarks and constitutes the 
bound (\ref{VI55l}), (\ref{boundonR}). 
The lower lines are upper bounds on $R_c$ and $R_b$, respectively, 
see (\ref{boundRheavy}). 
\label{fig:maxrat}}
\end{figure}
This dependence is shown in Fig.\ \ref{fig:maxrat} together 
with the bound (\ref{boundonR}) on $R$. Here we have used 
$m_c=1.3\,\mbox{GeV}$ and $m_b=4.6\,\mbox{GeV}$. 

We now confront the bound (\ref{boundonR}) with 
the available experimental data on $R$ at high energies 
which have been obtained by the NMC \cite{Arneodo:1996qe}, 
CCFR \cite{Yang:2001xc}, E143 \cite{Abe:1998ym}, 
EMC \cite{Aubert:1985fx} and CDHSW \cite{Berge:1989hr} 
collaborations. The scattering processes used to extract those 
data include not only $e^{\pm}p$ scattering but also 
other processes, among them muon and neutrino scattering 
on nuclear targets. In some of these processes one might in general 
expect additional caveats concerning the applicability of the 
dipole picture. For lack of further experimental data we 
nevertheless include the corresponding data points here. 
\begin{figure}[ht]
\begin{center}
\includegraphics[width=11.4cm]{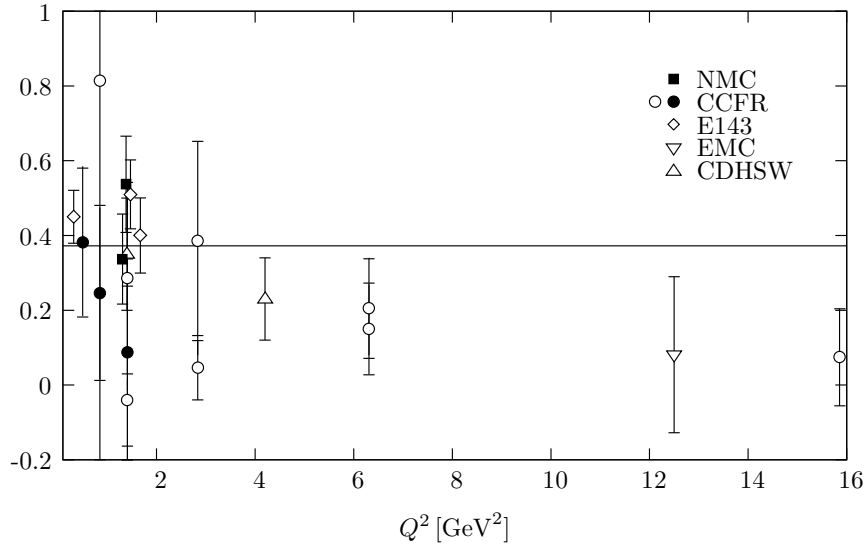}
\end{center}
\caption{Comparison of experimental data for $R=\sigma_L/\sigma_T$ in the 
region $x <0.05$ with the bound (\ref{boundonR}) resulting from 
the dipole picture. Full points correspond to data with $x <0.01$, 
open points are data with $0.01< x  \le 0.05$. 
\label{fig:bound}}
\end{figure}
In order to have sufficiently high energy we restrict ourselves 
to data points with $x < 0.05$. Fig.\ \ref{fig:bound} shows the 
available data together with the bound (\ref{boundonR}), 
where the few data points at $x< 0.01$ are represented by 
full points and those with $0.01 < x < 0.05$ as open points. 
The data have large errors, but by and large they 
respect the bound. For $Q^2$ below about $2\,\mbox{GeV}^2$, however, 
there seems to be the tendency that the data come close to the bound. 
Note the interesting fact that data very close to the bound could be 
accommodated in the dipole picture only if the dipole cross section 
$\hat{\sigma}^{(q)}$ were strongly peaked in $r$ around the maximum 
of $w_L^{(q)}/w_T^{(q)}$ -- which would appear to be very unlikely. 
Hence already data close to the bound can be interpreted as an 
indication of the breakdown of the dipole picture. In view of this 
we should be careful in interpreting the data at low $Q^2$ in terms 
of the dipole picture only. But the error bars 
of the presently available data on $R$ are clearly too large to draw firm 
conclusions about the range of applicability of the dipole picture 
at low $Q^2$. 

So far, no direct measurements of $F_L$ and $R$ have been done at HERA. 
A discussion of the available indirect determinations of $F_L$ in view 
of our bounds will be given elsewhere. Planned direct measurements of 
$F_L$ at HERA will hopefully lead to a better  determination of $R$, 
allowing for a stringent test of the validity of the dipole 
picture at low $Q^2$. 

A more detailed discussion of the bound on $R$  
is presented in the longer publication \cite{Ewerz:2006vd}, 
where we also discuss diffractive scattering and give examples 
for the behavior of $R$ in a typical saturation model. 

\boldmath
\section{Bounds on $F_2(W^2,Q_1^2)/F_2(W^2,Q_2^2)$} 
\unboldmath

Next we turn to the ratio of structure functions $F_2$ taken 
at the same energy $W$ but at different $Q^2$, for which 
we can derive bounds in a way similar to those on $R$. 
We recall that the dipole cross section $\hat{\sigma}^{(q)}$ 
depends on $r$ and $W^2$, and is independent of $Q^2$. 
(Note that this would not be the case if the energy variable 
in $\hat{\sigma}^{(q)}$ were $x$.) 
Therefore the cross sections of (\ref{sigTLdip}) and hence also 
$F_2$ involve the same factor $\hat{\sigma}^{(q)}(r,W^2)$ 
when evaluated at different $Q^2$. 
Then the non-negativity of the photon densities $w^{(q)}_{T,L}$ 
and of the dipole cross section $\hat{\sigma}^{(q)}$ 
implies for all $W^2$ the bounds 
\be
\label{boundonFdiffQ}
\frac{Q_1^2}{Q_2^2} \,\min_{q,r} 
\frac{w_T^{(q)}(r,Q_1^2) + w_L^{(q)}(r,Q_1^2)}
{w_T^{(q)}(r,Q_2^2) + w_L^{(q)}(r,Q_2^2)}
\, \le \,
\frac{F_2(W^2,Q_1^2)}{F_2(W^2,Q_2^2)}
\, \le \,
\frac{Q_1^2}{Q_2^2} \,\max_{q,r} 
\frac{w_T^{(q)}(r,Q_1^2) + w_L^{(q)}(r,Q_1^2)}
{w_T^{(q)}(r,Q_2^2) + w_L^{(q)}(r,Q_2^2)}
\,.
\ee
Also here, the most conservative bounds result from assuming the 
light quarks to be massless. 
\begin{figure}[ht]
\begin{center}
\includegraphics[width=11.4cm]{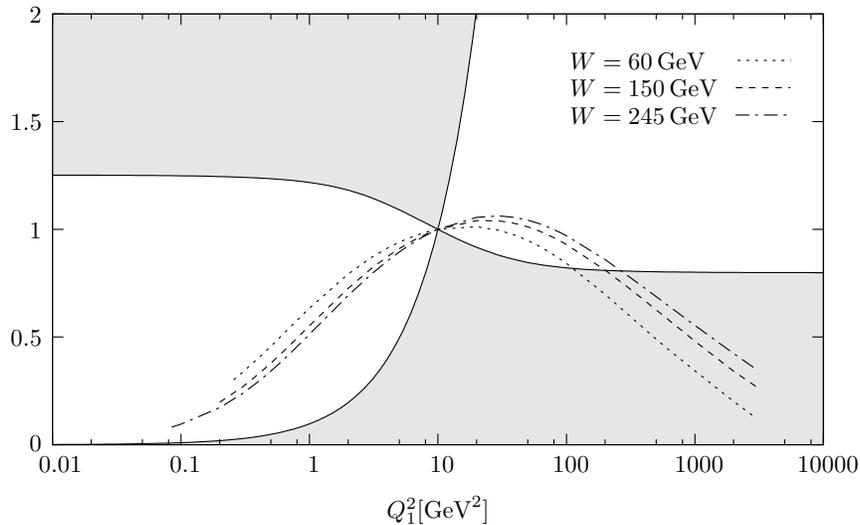}
\end{center}
\caption{The bounds (\ref{boundonFdiffQ}) on 
$F_2(W^2,Q_1^2)/F_2(W^2,Q_2^2)$ for $Q_2^2=10\,\mbox{GeV}^2$ 
and the corresponding fit to HERA data for three different 
values of $W$. Data in the shaded region cannot be described 
in the usual dipole picture. 
\label{fig:diffQminmax10}}
\end{figure}
They are shown in Fig.\ \ref{fig:diffQminmax10} 
for the choice $Q_2^2=10\,\mbox{GeV}^2$. 
The shaded region is excluded by the dipole picture. 

Turning to the corresponding data we first note that 
the systematic errors in measurements of $F_2$ at different 
$Q^2$ are certainly not independent, making a determination of 
the error on the ratio somewhat involved. Here we only want to 
illustrate the consequences of the bound and leave a more 
sophisticated treatment of the errors as well as the simultaneous 
variation of $Q_1^2$ and $Q_2^2$ for future work. 
We choose to compare the bound (\ref{boundonFdiffQ}) 
with the so-called ALLM fit to $F_2$ 
\cite{Abramowicz:1991xz,Abramowicz:1997ms} 
which represents the data within their errors, with the possible 
exception of very low $Q^2$ where it underestimates the 
data of \cite{Breitweg:2000yn}. 
We further fix $Q_2^2=10\,\mbox{GeV}^2$ and consider only 
three energies $W=60$, $150$, and $245\,\mbox{GeV}$, roughly covering 
the kinematical range of HERA. For a given $W$ we evaluate the 
ALLM fit only in the experimentally accessible $Q^2$ range. 
The $Q_1^2$-dependence of the ratios of $F_2$ thus computed is 
confronted with the bounds in Fig.\ \ref{fig:diffQminmax10}. 
The curves representing the data fall below the lower bound at 
values of $Q_1^2$ of about $100$ to $200\,\mbox{GeV}^2$, 
indicating a breakdown of the dipole picture. As expected, 
for larger $W$ the corresponding curve remains within the bounds 
up to a higher $Q_1^2$. The violation of the bound does not occur 
at a fixed $x$ though. For the energies $W=60$, $150$ and 
$245\,\mbox{GeV}$ the bound is violated at $x=0.03$, $0.008$ 
and $0.004$, respectively. 
At low $Q_1^2$ the ALLM fit does not come close to the bounds, 
hence its slight deviation from the data in this region is not relevant here. 

Again, we can also derive separate bounds for the heavy flavor 
contribution $F_2^{q \bar{q}}$, $q=c,b$, to $F_2$ which is obtained from 
$\sigma_{T,L}^{(q)}$ in analogy to (\ref{Ffromsigma}). 
These bounds then apply to 
$F^{q\bar{q}}_2(W^2,Q_1^2)/F^{q\bar{q}}_2(W^2,Q_2^2)$ and are 
of the same form as (\ref{boundonFdiffQ}), but with the minimum and 
maximum taken only over $r$ and not over $q$. 
\begin{figure}[ht]
\begin{center}
\includegraphics[width=11.4cm]{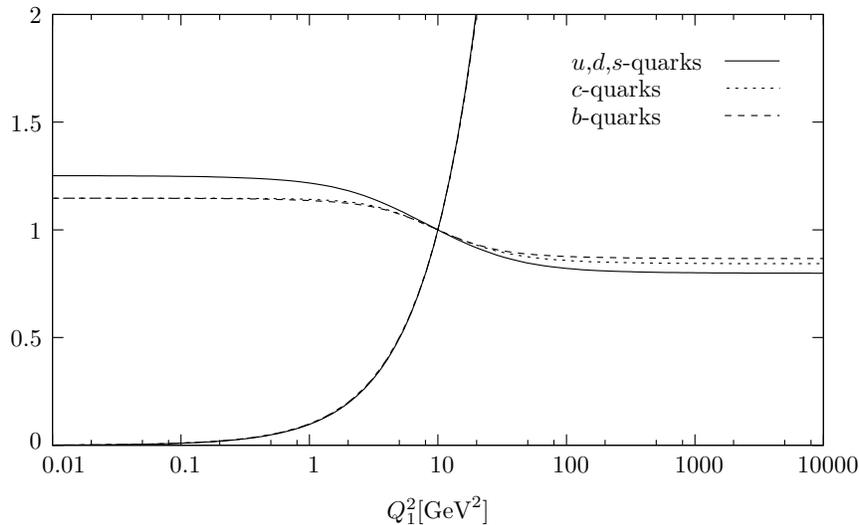}
\end{center}
\caption{The bounds on 
$F^{q\bar{q}}_2(W^2,Q_1^2)/F^{q\bar{q}}_2(W^2,Q_2^2)$ 
for $q=c,b$ together with the bounds (\ref{boundonFdiffQ}) 
(solid line) for $Q_2^2=10\,\mbox{GeV}^2$. 
\label{fig:diffQminmax10bc}}
\end{figure}
We show them together with the original bounds 
(\ref{boundonFdiffQ}) in Fig.\ \ref{fig:diffQminmax10bc}, 
again for $Q_2^2=10\,\mbox{GeV}^2$. 
For $Q_1^2 < Q_2^2$ the lower bound is insensitive to the 
quark mass, and the upper bound decreases with increasing 
quark mass. For $Q_1^2 > Q_2^2$, on the other hand, the 
upper bound is insensitive to the quark mass, and the lower 
bound increases with increasing quark mass. 

\section{Summary}

We have derived bounds on ratios of DIS nucleon structure functions 
from the usual color dipole picture. For this we have used the 
standard formulae for the photon wave functions and only the 
non-negativity of the dipole-nucleon cross sections. The bounds 
are hence independent of any model assumptions about the 
dipole-nucleon scattering. If a measurement of $R=\sigma_L/\sigma_T$ 
at given $W^2$ and $Q^2$ does not fulfill the bound 
(\ref{boundonR}) the dipole picture in the standard form is not 
valid there. If the ratio $F_2(W^2,Q_1^2)/F_2(W^2,Q_2^2)$ 
does not fulfill the bound (\ref{boundonFdiffQ}) the dipole 
picture in the standard form is not valid at least at one of the 
two kinematical points. In all these cases correction terms 
to the standard dipole picture as discussed in 
\cite{Ewerz:2004vf,Ewerz:2006vd} must play a significant role, 
or the dipole picture may even break down completely. 

In comparison with the data the bound on 
$R=\sigma_L/\sigma_T$ appears to suggest that we should 
be careful in interpreting the data at $Q^2 < 2\,\mbox{GeV}^2$ 
in terms of the dipole picture only. 
It is an interesting observation in this context that 
the possible evidence for saturation found by performing fits 
to the HERA data for $F_2$ in the dipole picture appears to 
depend on the inclusion of data at $Q^2 < 2\,\mbox{GeV}^2$ 
\cite{Forshaw:2004vv}, that is exactly in the region in which 
the data on $R$ come very close to the upper 
bound (\ref{boundonR}). For the bound on ratios of 
structure functions $F_2$ at the same $W^2$ but at different $Q^2$ 
the data constrain the use of the dipole picture for HERA energies 
to photon virtualities below around $100$ to $200\,\mbox{GeV}^2$. 
When analyzing HERA data in the framework of the dipole picture 
one should, therefore, restrict the analysis to the region allowed by 
these bounds in order to arrive at reliable conclusions. 

\section*{Acknowledgements}

We are grateful to J.\ Bartels, M.\ Klein, V.\ Lendermann, B.\ List, 
A.\ von Manteuffel, H.\,J.\ Pirner, and H.-C.\ Schultz-Coulon 
for helpful discussions. 
We thank A.\ Bodek and U.\,K.\ Yang for providing data for 
Fig.\ \ref{fig:bound}.


\begin{thebibliography}{99}

\bibitem{Breitweg:2000yn}
J.~Breitweg {\it et al.}  [ZEUS Collaboration],
Phys.\ Lett.\ B {\bf 487}, 53(2000) 
[arXiv:hep-ex/0005018].

\bibitem{Adloff:2000qk}
C.~Adloff {\it et al.}  [H1 Collaboration],
Eur.\ Phys.\ J.\ C {\bf 21}, 33 (2001) 
[arXiv:hep-ex/0012053].

\bibitem{Chekanov:2001qu}
S.~Chekanov {\it et al.}  [ZEUS Collaboration],
Eur.\ Phys.\ J.\ C {\bf 21}, 443 (2001) 
[arXiv:hep-ex/0105090].

\bibitem{Adloff:2003uh}
C.~Adloff {\it et al.}  [H1 Collaboration],
Eur.\ Phys.\ J.\ C {\bf 30}, 1 (2003) 
[arXiv:hep-ex/0304003].

\bibitem{Chekanov:2003yv}
S.~Chekanov {\it et al.}  [ZEUS Collaboration],
Phys.\ Rev.\ D {\bf 70}, 052001 (2004) 
[arXiv:hep-ex/0401003].

\bibitem{Gribov:1984tu}
L.~V.~Gribov, E.~M.~Levin and M.~G.~Ryskin,
Phys.\ Rept.\  {\bf 100}, 1 (1983).

\bibitem{Mueller:1985wy}
A.~H.~Mueller and J.~Qiu,
Nucl.\ Phys.\ B {\bf 268}, 427 (1986).

\bibitem{Mueller:1989st}
A.~H.~Mueller,
Nucl.\ Phys.\ B {\bf 335}, 115 (1990).

\bibitem{Golec-Biernat:1998js}
K.~Golec-Biernat and M.~W\"usthoff,
Phys.\ Rev.\ D {\bf 59}, 014017 (1999) 
[arXiv:hep-ph/9807513].

\bibitem{Golec-Biernat:1999qd}
K.~Golec-Biernat and M.~W\"usthoff,
Phys.\ Rev.\ D {\bf 60}, 114023 (1999) 
[arXiv:hep-ph/9903358].

\bibitem{Nikolaev:1990ja}
N.~N.~Nikolaev and B.~G.~Zakharov,
Z.\ Phys.\ C {\bf 49}, 607 (1991).

\bibitem{Nikolaev:et}
N.~N.~Nikolaev and B.~G.~Zakharov,
Z.\ Phys.\ C {\bf 53}, 331 (1992).

\bibitem{Mueller:1993rr}
A.~H.~Mueller,
Nucl.\ Phys.\ B {\bf 415}, 373 (1994).

\bibitem{Ewerz:2004vf}
C.~Ewerz and O.~Nachtmann,
arXiv:hep-ph/0404254.

\bibitem{Ewerz:2006vd}
C.~Ewerz and O.~Nachtmann,
arXiv:hep-ph/0604087.

\bibitem{Arneodo:1996qe}
M.~Arneodo {\it et al.}  [New Muon Collaboration],
Nucl.\ Phys.\ B {\bf 483}, 3 (1997) 
[arXiv:hep-ph/9610231].

\bibitem{Yang:2001xc}
U.~K.~Yang {\it et al.}  [CCFR/NuTeV Collaboration],
Phys.\ Rev.\ Lett.\  {\bf 87}, 251802 (2001) 
[arXiv:hep-ex/0104040].

\bibitem{Abe:1998ym}
K.~Abe {\it et al.}  [E143 Collaboration],
Phys.\ Lett.\ B {\bf 452}, 194 (1999) 
[arXiv:hep-ex/9808028].

\bibitem{Aubert:1985fx}
J.~J.~Aubert {\it et al.}  [European Muon Collaboration],
Nucl.\ Phys.\ B {\bf 259}, 189 (1985).

\bibitem{Berge:1989hr}
J.~P.~Berge {\it et al.},
Z.\ Phys.\ C {\bf 49}, 187 (1991).

\bibitem{Abramowicz:1991xz}
H.~Abramowicz, E.~M.~Levin, A.~Levy and U.~Maor,
Phys.\ Lett.\ B {\bf 269}, 465 (1991).

\bibitem{Abramowicz:1997ms}
H.~Abramowicz and A.~Levy,
arXiv:hep-ph/9712415.

\bibitem{Forshaw:2004vv}
J.~R.~Forshaw and G.~Shaw,
JHEP {\bf 0412}, 052 (2004)
[arXiv:hep-ph/0411337].

\end{thebibliography}
\end{document}